# Critique of the Fox-Lu model for Hodgkin-Huxley fluctuations in neuron ion channels

by

Ronald F. Fox
Smyrna, Georgia
July 20, 2018

"Subtle is the Lord, but malicious He is not."
Albert Einstein, Princeton, New Jersey, 1921

## Abstract

In 1976, Neher and Sakmann [1] published their pioneering work using the patch-clamp technique to measure ion currents through individual ion channels. In 1992 DeFelice and Isaac [2] published their analysis of global coupling of many channels across large domains of the cell membrane. They introduced an automaton model (Monte Carlo model, MC) that treated each and every channel subunit explicitly. After a seminar by DeFelice at Georgia Tech in 1992, Fox and Lu [3] showed that the MC model could be viewed as a master equation that allowed contraction into a Fokker-Planck equation, FP. Unfortunately such an equation is a partial differential equation and is difficult to simulate numerically. Using a well-known result that every FP equation has an antecedent Langevin equation, LE, Fox and Lu proposed such a description for ion channels in 1994. Their contraction followed the works of van Kampen [4] and of T. Kurtz [5]. The contraction produces a diffusion term with a state dependent diffusion matrix, **D**, that arises from the coupling matrix, **S**, in the LE. This **S** connected the noise terms to the channel subunit variables in the LE. Fox and Lu and many others later on observed that **SS** = **D**. Since **D** was determined by the contraction of the MC equations into the FP equation, this left the problem of determining the square root matrix, **S**, for every time step of the simulation. Since this is time consuming, Fox and Lu introduced simplified models not requiring the square root of a matrix. Subsequently, numerous studies were published that showed the several shortcomings of these simplified models. In 2011, Goldwyn et al. [6] rediscovered the overlooked original matrix dependent approach in the Fox-Lu 1994 paper.

They showed that it produced results in very good agreement with the MC results (algorithms for dealing with matrix square roots had improved during the time span 1994 - 2011). Something subtle had been overlooked, however. In 1991, Fox and Keizer [7] wrote a paper on an unrelated topic that utilized the work of van Kampen and of Kurtz. In that work the connection between **D** and **S** is $\mathbf{SS^T = D}$. $\mathbf{S^T}$ is the adjoint (transpose) of **S**. **D** remains a positive definite symmetric matrix but **S** need not be. After a fruitful correspondence with P. Orio in 2018, Fox has reproduced the 2012 results of Orio and Soudry [8] for potassium channels and has also found in closed form the solution for the more complicated sodium channels. All of this is done in the original Fox-Lu context with the $\mathbf{SS^T = D}$ correction to $\mathbf{SS = D}$. The square root problem generally must be done numerically, but the Cholesky (French mathematician, 1875-1918) factorization (or decomposition), as it is called, is always doable in closed form. Thereby the **S** matrix for sodium is given explicitly for the first time in this paper.

I. Introduction

The reader is referred to several papers in which the structures for the **D** matrices for potassium and sodium appear explicitly [3, 6, 8, 9]. These matrices are always positive definite and symmetric. The square root of a matrix with the properties of **D** is not unique. For example the matrix

$$\begin{bmatrix} 2 & 1 \\ 1 & 2 \end{bmatrix}$$

has two distinct square roots:

$$\begin{bmatrix} \dfrac{1}{2\sqrt{1-\dfrac{\sqrt{3}}{2}}} & \sqrt{1-\dfrac{\sqrt{3}}{2}} \\ \sqrt{1-\dfrac{\sqrt{3}}{2}} & \dfrac{1}{2\sqrt{1-\dfrac{\sqrt{3}}{2}}} \end{bmatrix} \quad \text{and} \quad \begin{bmatrix} \dfrac{1}{2\sqrt{1+\dfrac{\sqrt{3}}{2}}} & \sqrt{1+\dfrac{\sqrt{3}}{2}} \\ \sqrt{1+\dfrac{\sqrt{3}}{2}} & \dfrac{1}{2\sqrt{1+\dfrac{\sqrt{3}}{2}}} \end{bmatrix}$$

Larger matrices have more square roots. Generally an n × n positive definite symmetric matrix will have $2^n$ distinct square roots. These multiple roots correspond to the ± square roots of the diagonal elements of the diagonalized matrix. If we always choose the positive square roots then a positive definite symmetric matrix has a unique positive definite symmetric square root. This fact was used in all of the papers based on the Fox-Lu method. Consequently the result for **S** turned out to satisfy the additional property $\mathbf{S}^T = \mathbf{S}$. Thus the distinction between $\mathbf{SS} = \mathbf{D}$ and $\mathbf{SS}^T = \mathbf{D}$ appeared not to matter.

An n variable Langevin equation (LE) has the form

$$\frac{d}{dt}\vec{x} = \vec{f}(\vec{x}) + \mathbf{S}(\vec{x}, t) \cdot \tilde{\vec{g}}(t)$$

in which $\vec{x}$ is an n-component vector, $\vec{f}(\vec{x})$ is an n-component force and $\mathbf{S}(\vec{x}, t)$ has an n × n matrix representation. The n components of $\tilde{\vec{g}}(t)$ are *independent* Gaussian white noise terms with zero mean and covariances equal to 2. If we prefer to write the LE with explicit indices then the equation takes the form

$$\frac{d}{dt}\vec{x}_i = \vec{f}_i(\vec{x}) + S_{ij}(\vec{x}, t)\tilde{g}_j(t)$$

wherein the repeated index $j$ is summed over from 1 to n. There is a direct way to express the associated Fokker-Planck equation (FP) using time ordered operator cumulants (van Kampen [10], Fox [11]). Let $\rho(\vec{x}, t)$ denote the probability density function for $\vec{x}$. It satisfies the probability continuity equation

$$\frac{\partial}{\partial t}\rho(\vec{x}, t) = -\frac{\partial}{\partial x_i}\left(\left(\frac{d}{dt}x_i\right)\rho(\vec{x}, t)\right)$$

$$= -\frac{\partial}{\partial x_i}\left((\vec{f}_i(\vec{x}) + S_{ij}(\vec{x}, t)\tilde{g}_j(t))\rho(\vec{x}, t)\right)$$

This equation has a formal solution given by a time ordered exponential

$$\rho(\vec{x}, t) = \overleftarrow{T}exp\left[-\int_0^t ds \frac{\partial}{\partial x_i}(\vec{f}_i(\vec{x}) + S_{ij}(\vec{x}, t)\tilde{g}_j(t))\right]\rho(\vec{x}, 0)$$

The FP distribution, $P(\vec{x}, t)$, is the stochastic average of $\rho(\vec{x}, t)$. Since the noise terms are independent *white* noises the first and second operator cumulants are the only non-vanishing cumulants. The second cumulant generates the diffusion terms. The second cumulant is given by the second moment of the noise terms (the first moments vanish) and has the form

$$\frac{1}{2}\int_0^t\int_0^t ds'ds \, \langle \frac{\partial}{\partial x_i} S_{ij}(\vec{x}, s')\tilde{g}_j(s') \frac{\partial}{\partial x_k} S_{kl}(\vec{x}, s)\tilde{g}_l(s) \rangle$$

where the brackets, $\langle ... \rangle$ denote averaging. The assumed properties of the white noises imply

$$\langle \tilde{g}_j(s')\tilde{g}_l(s) \rangle = 2\delta_{jl}\delta(s' - s)$$

Putting this into the preceding equation yields

$$t \frac{\partial^2}{\partial x_i \partial x_k} S_{ij}(\vec{x}, s') \, S_{kj}(\vec{x}, s')$$

A careful noting of the order of the indices shows that this contains $\mathbf{SS^T}$ and not $\mathbf{SS}$. Therefore the diffusion matrix is $\mathbf{D = SS^T}$, or equivalently

$$D_{ik} = S_{ij}S^T_{jk}$$

Given $\mathbf{D}$ we seek $\mathbf{S}$. For positive definite symmetric $\mathbf{D}$ there is a *unique* Cholesky decomposition of the form of $\mathbf{SS^T}$ with the additional property that $\mathbf{S}$ is lower (or upper) *triangular* with non-zero elements along the diagonal and any square roots are chosen with the + sign. Nothing could be more different compared to $\mathbf{S = S^T}$.

II. Potassium channels

The potassium channels contain 4 identical n-type subunits. The subunits may be in an open state or in a closed state. All four subunits need to be open for conductance of a potassium ion. If $x_i$, ($x_0, x_1, x_2, x_3$ and $x_4$), represents the fraction of channels in area A with an areal density $N_K$ that have $i$ open subunits then these 5 variables sum to 1. Instead of treating the problem with 5 variables we choose to use 4 so that the size the matrices involved is smaller. Later we do the same for sodium. In Orio and Soudry [8] the larger number is used, and in Goldwyn et al. [6] the smaller number is used. Thus we have the result for 4 independent variables that corresponds with the result for 5 constrained variables, and for the sodium case the first explicit results for 7 variables.

The **D** matrix for Potassium is

$$D_K = \frac{1}{N_K} \begin{bmatrix} d_1 & d_{12} & 0 & 0 \\ d_{12} & d_2 & d_{23} & 0 \\ 0 & d_{23} & d_3 & d_{34} \\ 0 & 0 & d_{34} & d_4 \end{bmatrix}$$

in which the non-zero matrix elements are given by

$$d_1 = 4\alpha_n x_0 + (3\alpha_n + \beta_n)x_1 + 2\beta_n x_2$$
$$d_2 = 3\alpha_n x_1 + 2(\alpha_n + \beta_n)x_2 + 3\beta_n x_3$$
$$d_3 = 2\alpha_n x_2 + (\alpha_n + 3\beta_n)x_3 + 4\beta_n x_4$$
$$d_4 = \alpha_n x_3 + 4\beta_n x_4$$
$$d_{12} = -(3\alpha_n x_1 + 2\beta_n x_2)$$
$$d_{23} = -(2\alpha_n x_2 + 3\beta_n x_3)$$
$$d_{34} = -(\alpha_n x_3 + 4\beta_n x_4)$$
$$x_0 \equiv 1 - x_1 - x_2 - x_3 - x_4$$

Here the $x_i$'s are taken to be their instantaneous values rather than their steady state values as is done in [6, 8] although the steady state approximation may also be applied if desired. The $x_i$'s are dimensionless, the $\alpha_n$ and $\beta_n$ rates have dimension sec$^{-1}$, and N$_K$ has dimension cm$^{-2}$.

The Cholesky decomposition is expressed as

$$\begin{bmatrix} d_1 & d_{12} & 0 & 0 \\ d_{12} & d_2 & d_{23} & 0 \\ 0 & d_{23} & d_3 & d_{34} \\ 0 & 0 & d_{34} & d_4 \end{bmatrix} = \begin{bmatrix} s_1 & 0 & 0 & 0 \\ s_{12} & s_2 & 0 & 0 \\ s_{13} & s_{23} & s_3 & 0 \\ s_{14} & s_{24} & s_{34} & s_4 \end{bmatrix} \times \begin{bmatrix} s_1 & s_{12} & s_{13} & s_{14} \\ 0 & s_2 & s_{23} & s_{24} \\ 0 & 0 & s_3 & s_{34} \\ 0 & 0 & 0 & s_4 \end{bmatrix}$$

This system of equations is solved by starting in the upper left-hand corner and working down the rows, moving left to right along a row, until done. Triangularity is the key.

A shorthand notation proves useful. Let the following combinations that are equal to sub-block determinants in **D** be given by

$$\Delta_{12} \equiv d_1 d_2 - d_{12}^2$$
$$\Delta_{123} \equiv d_1 d_2 d_3 - d_{12}^2 d_3 - d_{23}^2 d_1$$
$$\Delta_{1234} \equiv d_1 d_2 d_3 d_4 - d_{12}^2 d_3 d_4 - d_{23}^2 d_1 d_4 - d_{34}^2 d_1 d_2 + d_{34}^2 d_{12}^2$$

The Cholesky matrix elements work out to be

$$s_1 = \sqrt{d_1}$$
$$s_2 = \frac{\sqrt{\Delta_{12}}}{\sqrt{d_1}}$$
$$s_3 = \frac{\sqrt{\Delta_{123}}}{\sqrt{\Delta_{12}}}$$
$$s_4 = \frac{\sqrt{\Delta_{1234}}}{\sqrt{\Delta_{123}}}$$
$$s_{12} = \frac{d_{12}}{\sqrt{d_1}}$$
$$s_{23} = \frac{\sqrt{d_1} d_{23}}{\sqrt{\Delta_{12}}}$$
$$s_{34} = \frac{\sqrt{\Delta_{12}} d_{34}}{\sqrt{\Delta_{123}}}$$
$$s_{13} = 0$$
$$s_{24} = 0$$
$$s_{14} = 0$$

### III. Sodium channels

The sodium channels contain 3 identical m-type subunits and 1 h-type subunit. The subunits may be in an open state or in a closed state. All four subunits need to be open for conductance of a sodium ion. Let $y_{ij}$ represent the fraction of channels in area A with an areal density $N_{Na}$ that have $i$ open m-type subunits and $j$ open h-type subunits ($i = 0, 1, 2, 3;\ j = 0, 1$). These 8 quantities sum to 1.

The **D** matrix for sodium is

$$D_{Na} = \frac{1}{N_{Na}} \begin{bmatrix} d_1 & d_{12} & 0 & 0 & d_{15} & 0 & 0 \\ d_{12} & d_2 & d_{23} & 0 & 0 & d_{26} & 0 \\ 0 & d_{23} & d_3 & 0 & 0 & 0 & d_{35} \\ 0 & 0 & 0 & d_4 & d_{45} & 0 & 0 \\ d_{15} & 0 & 0 & d_{45} & d_5 & d_{56} & 0 \\ 0 & d_{26} & 0 & 0 & d_{56} & d_6 & d_{67} \\ 0 & 0 & d_{35} & 0 & 0 & d_{67} & d_7 \end{bmatrix}$$

in which the non-zero matrix elements are given by

$$d_1 = 3\alpha_m y_{00} + (2\alpha_m + \beta_m + \alpha_h)y_{10} + 2\beta_m y_{20} + \beta_h y_{11}$$
$$d_2 = 2\alpha_m y_{10} + (\alpha_m + 2\beta_m + \alpha_h)y_{20} + 3\beta_m y_{30} + \beta_h y_{21}$$
$$d_3 = \alpha_m y_{20} + (3\beta_m + \alpha_h)y_{30} + \beta_h y_{31}$$
$$d_4 = \alpha_h y_{00} + (3\alpha_m + \beta_h)y_{01} + \beta_m y_{11}$$
$$d_5 = \alpha_h y_{10} + 3\alpha_m y_{01} + (2\alpha_m + \beta_m + \beta_h)y_{11} + 2\beta_m y_{21}$$
$$d_6 = \alpha_h y_{20} + 2\alpha_m y_{11} + (\alpha_m + 2\beta_m + \beta_h)y_{21} + 3\beta_m y_{31}$$
$$d_7 = \alpha_h y_{30} + \alpha_m y_{21} + (3\beta_m + \beta_h)y_{31}$$
$$d_{12} = -2(\alpha_m y_{10} + \beta_m y_{20})$$
$$d_{15} = -(\alpha_h y_{10} + \beta_h y_{11})$$
$$d_{23} = -(\alpha_m y_{20} + 3\beta_m y_{30})$$
$$d_{26} = -(\alpha_h y_{20} + \beta_h y_{21})$$
$$d_{37} = -(\alpha_h y_{30} + \beta_h y_{31})$$
$$d_{45} = -(3\alpha_m y_{01} + \beta_m y_{11})$$
$$d_{56} = -2(\alpha_m y_{11} + \beta_m y_{21})$$
$$d_{67} = -(\alpha_m y_{21} + 3\beta_m y_{31})$$
$$y_{00} \equiv 1 - y_{01} - y_{10} - y_{20} - y_{30} - y_{11} - y_{21} - y_{31}$$

Here the $y_{ij}$'s are taken to be their instantaneous values rather than their steady state values as is done in [6, 8] although the steady state approximation may also be applied if desired.

The Cholesky factorization can be expressed in parallel with the potassium case. We have the $D_{Na}$ matrix above and if we leave off the density factor $1/N_{Na}$ the remainder must be equal to $SS^T$ where $S$ is given by

$$\begin{bmatrix} s_1 & 0 & 0 & 0 & 0 & 0 & 0 \\ s_{12} & s_2 & 0 & 0 & 0 & 0 & 0 \\ s_{13} & s_{23} & s_3 & 0 & 0 & 0 & 0 \\ s_{14} & s_{24} & s_{34} & s_4 & 0 & 0 & 0 \\ s_{15} & s_{25} & s_{35} & s_{45} & s_5 & 0 & 0 \\ s_{16} & s_{26} & s_{36} & s_{46} & s_{56} & s_6 & 0 \\ s_{17} & s_{27} & s_{37} & s_{47} & s_{57} & s_{67} & s_7 \end{bmatrix}$$

As with the potassium case, begin multiplying $S$ by $S^T$ in the upper left hand corner and proceed down the rows, moving from left to right along a row, solving for the $S$ matrix elements in terms of those for $D$ until done.

While this process is as straightforward for sodium as it was for potassium it is considerably more complicated. To help several shorthand notations are introduced. These are sub-block determinants emanating from $D$.

$$\Delta_{12} \equiv d_1 d_2 - d_{12}^2$$
$$\Delta_{123} \equiv d_1 d_2 d_3 - d_{12}^2 d_3 - d_{23}^2 d_1$$
$$\Delta_{23} \equiv d_2 d_3 - d_{23}^2$$
$$\Delta_{45} \equiv d_4 d_5 - d_{45}^2$$
$$\Delta_{456} \equiv d_4 d_5 d_6 - d_{45}^2 d_6 - d_{56}^2 d_4$$

The Cholesky matrix elements work out to be:

$$s_1 = \sqrt{d_1}$$
$$s_2 = \frac{\sqrt{\Delta_{12}}}{\sqrt{d_1}}$$
$$s_3 = \frac{\sqrt{\Delta_{123}}}{\sqrt{\Delta_{12}}}$$

$$s_4 = \sqrt{d_4} = \frac{\sqrt{d_4 \Delta_{123}}}{\sqrt{\Delta_{123}}}$$

$$s_5 = \sqrt{\left(\frac{\Delta_{45}}{d_4} - d_{15}^2 \frac{\Delta_{23}}{\Delta_{123}}\right)} = \sqrt{\frac{\Delta_{45}\Delta_{123} - d_{15}^2 d_4 \Delta_{23}}{d_4 \Delta_{123}}}$$

The right-most expressions are in terms of the sub-block determinants. If we label the $n \times n$ block determinant by its size, $n$, then calling it $\Delta_n$ gives

$$\Delta_1 = d_1$$
$$\Delta_2 = d_1 d_2 - d_{12}^2 = \Delta_{12}$$
$$\Delta_3 = d_1 d_2 d_3 - d_{12}^2 d_3 - d_{23}^2 d_1 = \Delta_{123}$$
$$\Delta_4 = d_4(d_1 d_2 d_3 - d_{12}^2 d_3 - d_{23}^2 d_1) = d_4 \Delta_{123}$$
$$\Delta_5 = \Delta_{45}\Delta_{123} - d_{15}^2 d_4 \Delta_{23}$$
$$\Delta_6 = (\Delta_{456}\Delta_{123} - d_1 d_3 d_{26}^2 \Delta_{45} - d_4 d_6 d_{15}^2 \Delta_{23} - 2 d_4 d_{56} d_{12} d_{15} d_{26} d_3$$
$$\qquad + d_3 d_4 d_{15}^2 d_{26}^2)$$

$$s_5 = \frac{\sqrt{\Delta_5}}{\sqrt{\Delta_4}}$$

$$s_6 = \frac{\sqrt{\Delta_6}}{\sqrt{\Delta_5}}$$

$$s_7 = \frac{\sqrt{\Delta_7}}{\sqrt{\Delta_6}}$$

in which $\Delta_7$ appears. We have computed this determinant of the full $7 \times 7$ **D** matrix and arrived at 44 distinct terms. While this is rather few in number compared to 7!, because of the abundance of zeros in **D**, it is not the most efficient presentation for computation. Instead one can use the equation producing $s_7$ during the $\mathbf{SS}^T$ multiplication. This means use

$$s_7 = \sqrt{d_7 - s_{37}^2 - s_{57}^2 - s_{67}^2}$$

in which the off-diagonal matrix elements are given below. They are produced by the computation before they are needed in $s_7$. Continuing with the solutions we get

$$s_{12} = \frac{d_{12}}{\sqrt{d_1}}$$
$$s_{13} = 0$$
$$s_{14} = 0$$
$$s_{15} = \frac{d_{15}}{\sqrt{d_1}}$$
$$s_{16} = 0$$
$$s_{17} = 0$$
$$s_{23} = \frac{d_{23}\sqrt{d_1}}{\sqrt{\Delta_{12}}}$$
$$s_{24} = 0$$
$$s_{25} = -\frac{d_{12}d_{15}}{\sqrt{d_1 \Delta_{12}}}$$
$$s_{26} = \frac{d_{26}\sqrt{d_1}}{\sqrt{\Delta_{12}}}$$
$$s_{27} = 0$$
$$s_{34} = 0$$
$$s_{35} = \frac{d_{23}d_{12}d_{15}}{\sqrt{\Delta_{12}\Delta_{123}}}$$
$$s_{36} = -\frac{d_1 d_{23} d_{26}}{\sqrt{\Delta_{12}\Delta_{123}}}$$
$$s_{37} = \frac{d_{37}\sqrt{\Delta_{12}}}{\sqrt{\Delta_{123}}}$$
$$s_{45} = \frac{d_{45}}{\sqrt{d_4}}$$
$$s_{46} = 0$$
$$s_{47} = 0$$
$$s_{56} = \frac{\sqrt{\Delta_4}}{\sqrt{\Delta_5}}\left(d_{56} + \frac{d_3 d_{12} d_{15} d_{26}}{\Delta_{123}}\right)$$
$$s_{57} = -\frac{d_{12}d_{15}d_{23}d_{37}}{\Delta_{123}}\frac{\sqrt{\Delta_4}}{\sqrt{\Delta_5}}$$

$$s_{67} = \left( d_{67} + \frac{d_1 d_{23} d_{26} d_{37}}{\Delta_{123}} \right.$$
$$\left. + \frac{(d_{56}\Delta_{123} + d_3 d_{12} d_{15} d_{26}) d_{12} d_{15} d_{23} d_{37}}{\Delta_{123}^2} \frac{\Delta_4}{\Delta_5} \right) \frac{\sqrt{\Delta_5}}{\sqrt{\Delta_6}}$$

The diagonal matrix elements were given first in this tabulation but occur naturally only when their row is started. That way off-diagonal elements occur before they are needed in the diagonal terms.

## IV. Commentary

The result of a contraction of the description starting with a master equation is a Fokker-Planck equation with a diffusion term that contains a state dependent diffusion matrix, **D**. For the DeFelice-Isaac master equation (MC model) the elements of the diffusion matrix depend on the instantaneous values of the membrane voltage, V, and of the channel subunit parameters $x_i$ or $y_{ij}$. Because the Fokker-Planck equation is a partial differential equation numerical implementation is demanding. Consequently an equivalent Langevin equation is sought for numerical simulations so that it's corresponding Fokker-Planck equation is the same as the one derived from the master equation. This is a kind of inverse problem. Apparently it has more than one solution. In the original Fox-Lu model the connection between the Langevin **S** matrix and the Fokker-Planck **D** matrix is $\mathbf{SS}^\mathbf{T} = \mathbf{D}$. However if one diagonalizes **D** and then takes the square roots of the diagonal elements and calls this the transformed **S** one can reverse the diagonalizing transformation to get back **D** and an **S** that satisfies $\mathbf{SS} = \mathbf{D}$, or $\mathbf{S} = \mathbf{S}^\mathbf{T}$. This is a special case for **S** in which **S** is the square root of **D**. Getting **S** at each integration time step is labor intensive. Because we still get back the original Fokker-Planck equation the statistics generated by the Langevin equation using the **S** square root are very good when compared with the master equation output.

An alternative Langevin equation is possible by interpreting the relation $\mathbf{SS}^\mathbf{T} = \mathbf{D}$ to mean find the Cholesky solution for **S**. In this paper we have shown how to do this in closed form for both the potassium and the sodium channels. The potassium results agree with those found by Orio and Soudry (they use 5 variables and we use 4) and the 7-variable sodium results are new.

A purely mathematical puzzle shows itself in these results. Sub-block determinants of the **D** matrices play a major role in the structure of the **S** matrix elements. This suggests that a universal form for **S** may exist. It also explains why the positive definiteness of **D** implies that the arguments of the square roots in these formulas are positive.

# Appendix

After noting that there could be closed form expressions for the Cholesky decomposition of a diffusion matrix, such expressions were derived. A diffusion matrix, **D**, for a Fokker-Planck equation is a symmetric, positive-definite matrix. The Cholesky decomposition matrix, **S**, satisfies $SS^T = D$. **S** is lower triangular. Elementary properties of determinants imply

$$det\, S = det\, S^T$$

$$det\, S = \sqrt{det\, D}$$

$$det\, S = \prod_{j=1}^{n} S_j$$

In the third equality the $S_j$'s are the diagonal elements of a $n \times n$ **S** matrix. The result is due to triangularity of **S**.

Let the following definitions be introduced to simplify notation:

$$\Delta_0 \equiv 1, \quad \Delta_1 \equiv d_1 \quad \Delta_2 \equiv d_1 d_2 - d_{12}^2$$

$$\Delta_3 = d_1 d_2 d_3 + 2 d_{12} d_{13} d_{23} - d_1 d_{23}^2 - d_2 d_{13}^2 - d_3 d_{12}^2$$

Starting in the upper left-hand corner of **D**, form the sub-matrices of sizes 1×1, 2×2, 3×3, …, $n \times n$. The deltas are the determinants of these sub-matrices with subscripts up to $n$ and $\Delta_0$ is defined for convenience.

It is straightforward to deduce the diagonal element formula

$$S_j = \frac{\sqrt{\Delta_j}}{\sqrt{\Delta_{j-1}}}$$

for $j = 1, 2, 3, \ldots, n$. See the three identities at the start of this appendix for verification.

Of the remaining off-diagonal elements of **S** there are just two types. There are the elements next to the diagonal with indices $S_{jj+1}$. And there are the elements with indices $S_{jk}$ for $k = j + 2, j + 3, \ldots, n$. Together these elements determine the entire non-zero part of row $j$ of $\mathbf{S}^\mathrm{T}$.

To understand the formula for $S_{jj+1}$ consider the $j + 1 \times j + 1$ **D** sub-matrix (from the upper left-hand corner) and form the cofactor of element $d_{jj+1}$. Recall that the cofactor is the determinant of the $j + 1 \times j + 1$ sub-matrix after the row $j$ and the column $j + 1$ have been removed. There is also a factor of $(-1)^{2j+1}$ that is always $-1$ for us. With this definition the result for $S_{jj+1}$ works out to be

$$S_{jj+1} = -\frac{\mathrm{cofactor}(d_{jj+1})}{\sqrt{\Delta_{j-1}\Delta_j}}$$

The formula for $S_{jk}$ for $k = j + 2, j + 3, \ldots, n$ is identical in form with the expression above with the change that every occurrence of $j + 1$ is replaced by $k$.

## References

...